\documentclass[9pt,5p,twocolumn]{elsarticle}
\usepackage{graphicx,dcolumn,bm,amsmath,amssymb,mathrsfs}
\usepackage{lineno,hyperref}
\modulolinenumbers[5]

\journal{PLA, and published as: M.~Cong, C.~Li, B.-Q.~Ma, Phys. Lett. A 383 (2019) 1836-1844}

\begin{document}

\newcommand{\cdotn}{\!\cdot\!}
\newcommand{\dd}{\mathrm{d}}
\newcommand{\abs}[1]{\big| #1 \big|}

\begin{frontmatter}

\title{First digit law from Laplace transform}

\author[pku,hku]{Mingshu Cong}
\author[pku]{Congqiao Li}
\author[pku,CICQM,CHEP]{Bo-Qiang Ma\corref{ma}}
\cortext[ma]{Corresponding author}
\ead{mabq@pku.edu.cn}

\address[pku]{School of Physics and State Key Laboratory of Nuclear Physics and Technology, \\Peking University, Beijing 100871, China}
\address[CICQM]{Collaborative Innovation Center of Quantum Matter, Beijing, China}
\address[CHEP]{Center for High Energy Physics, Peking University, Beijing 100871, China}
\address[hku]{The FinTech and Blockchain Laboratory, Department of Computer Science, The University of Hong Kong, Hong Kong, China}

\begin{abstract}

The occurrence of digits 1 through 9 as the leftmost nonzero
digit of numbers from real-world sources is distributed unevenly
according to an empirical law, known as Benford's law or the first
digit law. It remains obscure why a variety of data sets generated from
quite different dynamics obey this particular law.
We perform a study of Benford's law from the application of the Laplace transform, and find that the logarithmic Laplace spectrum of the digital indicator
function can be approximately taken as a constant. This particular constant, being exactly
the Benford term, explains the prevalence of Benford's law. The slight variation
from the Benford term leads to deviations from Benford's law for distributions
which oscillate violently in the inverse Laplace space.
We prove that the whole family of completely monotonic
distributions can satisfy Benford's law within a small bound.
Our study suggests that Benford's law originates from the way
that we write numbers, thus should be taken as a basic mathematical knowledge.

\end{abstract}


\end{frontmatter}

%\linenumbers

\section{Introduction}
\begin{figure}[h]
	\begin{center}
		\includegraphics[width=.7\linewidth]{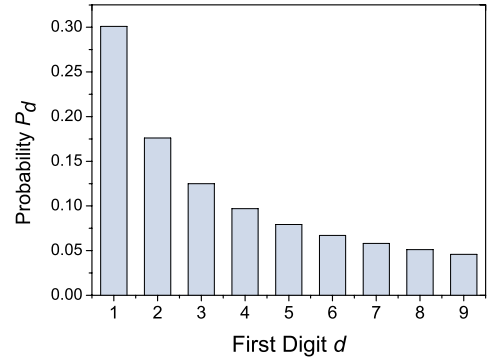}
		\caption{Benford's law of the first digit distribution, from which we see that the probability of finding numbers with leading digit 1 is more than 6 times larger than that with 9.}\label{Benfordfigure}
	\end{center}
\end{figure}

There is an empirical law concerning the occurrence of the first digits in real-world data, stating that the first digits of natural numbers prefer small ones rather than a uniform distribution as might be expected. More accurately, the probability that a number begins with digit $d$, where $d=1,2,\dots,9$ respectively, can be expressed as
\begin{equation}
\label{benford}
P_d = \log_{10}(1+\frac{1}{d})\,,\quad  d=1, 2,\dots, 9\,,
\end{equation}
as shown in Fig.~\ref{Benfordfigure}. This is known as Benford's law, which is also called the first digit law or the significant digit law, first noticed by Newcomb~\cite{n81} in 1881, and then re-discovered independently by Benford~\cite{b38} in 1938.

Empirically, %the populations of countries,
the areas of lakes, the lengths of rivers, the Arabic numbers on the front page of a newspaper~\cite{b38}, physical constants~\cite{bk91}, the stock market indices~\cite{l96}, file sizes in a personal computer~\cite{tfgs07}, survival distributions~\cite{lse00}, etc., %{\it et al.},
all conform to this peculiar law well. Due to the powerful data analyzing tools provided by computer science, Benford's law has been verified for a vast number of examples in various domains, such as economics~\cite{giles2007benford,nigrini2015persistent}, social science~\cite{lse00}, environmental science~\cite{b05}, biology~\cite{biology}, geology~\cite{geology}, astronomy~\cite{sm10a}, statistical physics~\cite{sm10b,sm10c}, nuclear physics~\cite{nr08,liu2011benford,hui2011benford}, particle physics~\cite{sm09a}, and some dynamical systems~\cite{bbh05,b05dcds}. There have been also many explorations on the applications of the law in various fields, e.g., in upgrading the description in precipitation regime shift~\cite{yang2017out}. Some applications focus on detecting data and judging their reasonableness, such as distinguishing and ascertaining fraud in taxing and accounting~\cite{n96,durtschi2004effective,tam2007breaking,nigrini2016implications}, fabrication in clinical trials~\cite{marzouki2005}, the authenticity of the pollutant concentrations in ambient air~\cite{b05}, electoral cheats or voting anomalies~\cite{tfgs07,leemann2014systematic}, and falsified data in scientific experiments~\cite{d07a}. Moreover, the first digit law is applied in computer science for speeding up calculation~\cite{barlow1985}, minimizing expected storage space~\cite{s88,bh07}, analyzing the behavior of floating-point arithmetic algorithms~\cite{bh07}, and also for various studies in the image domain~\cite{j01,f07}.

Theoretically, several elegant properties of Benford's law have been revealed. In mathematics, Benford's law is the only digit law that is scale-invariant~\cite{p61,bhm08}, which means that the law does not depend on any particular choice of units. This law is also base-invariant~\cite{h95a,h95b,h95c}, which means that it is independent of the base $b$. In the octal system ($ b=8 $), the hexadecimal system ($ b=16 $), or other base systems, the data, if fit the law in the decimal system ($b=10$), all fit the general Benford's law
\begin{equation}\label{benbase}
P_d = \log_{b}(1+\frac{1}{d})\,, \quad d=1, 2,\dots, {b-1}\,.
\end{equation}
The law is also found to be power-invariant~\cite{sm09a}, i.e., any power ($\neq 0$) on numbers in the data set does not change the first digit distribution.

There have been many studies on Benford's law with numerous breakthroughs.
For example, Hill provided a measure-theoretical proof that Benford's law is equivalent to the scale-invariant property and that random samples taken from randomly-selected distributions converge to Benford's law~\cite{h95b,h95c}. Pietronero {\it et al.} explained why some data sets naturally show scale-invariant properties from a dynamics governed by multiplicative fluctuations thus conform to Benford's law~\cite{pietronero2001explaining}. Gottwald and Nicol figured out that deterministic quasiperiodic or periodic forced multiplicative process and even affine processes also tend to Benford's law~\cite{gottwald2002nature}. Engel and Leuenberger focused on exponential distributions and illustrated that they approximately obey Benford's law within a bound of 0.03~\cite{el03}. Smith applied digital signal processing and studied the distributions on the logarithmic scale and their frequency domain, revealing that the first digit law holds for distributions with no components of nonzero integer frequencies~\cite{smith1997scientist}. Fewster asserted that any distribution might tend to Benford's law if it can span several orders of magnitude and be reasonably smooth~\cite{fewster2009simple}.

However, there are still various data sets that violate Benford's
law, e.g., the telephone numbers, birthday data, and accounts with
a fixed minimum or maximum. Benford's law still remains obscure
whether this law is merely a result of our way of writing numbers.
If the answer is yes, why not all number sets obey this law; if
the answer is no, why is this law so common that it can be a good
approximation for most data sets. The situation can also be
reflected by some puzzles about Benford's law in the literature,
e.g., it is stated by Tao that no one can really prove or derive
this law because Benford's law, being an empirically observed
phenomenon rather than an abstract mathematical fact, cannot be
``proved'' the same way a mathematical theorem can be
proved~\cite{tao2010epsilon}. Aldous and Phan also suggested that
without checking the assumptions of Benford's law for the data
sets we studied, this logically correct mathematical theorem is
not relevant to the real world~\cite{aldous2010can}.

%In fact, Berger and Hill also insisted that there is no simple explanation for Benford's law~\cite{berger2011basic}.

Therefore, most studies on Benford's law are case studies in
literature, restricted to a specific probability density
distribution or a group of them. In this work, we provide a
general derivation of Benford's law with the application of the
Laplace transform, which is an important tool of mathematical
methods in physics~\cite{mathphy}. From our derivation, we can
safely assert that the deviation from Benford's law is always less
than a small proportion of the $ L^1 $-norm of the logarithmic inverse Laplace
transform of the probability density function. This bound is
universal. Since the $ L^1 $-norm of the logarithmic inverse Laplace transform
is usually small but not zero, Benford's law is commonly well
obeyed but not strictly obeyed. We introduce a guideline to judge
how well a specific distribution obeys Benford's law. In this
method, the degree of deviation from the law is associated with
the oscillatory behavior of the probability density function in
the inverse Laplace space. We find that the whole family of completely
monotonic distributions can all fulfill Benford's law within a small bound.
We also carry out some numerical
estimations of the error term, and present several examples which
verify our method. We agree with Goudsmit and
Furry~\cite{goudsmit} and reveal from our own method that the
appearance of the first digit law is a logical consequence of the
digital system, but not due to some unknown mechanics of the
nature.

Our work is organized as follows.
In Sec.~\ref{mantissa} we introduce the digital indicator functions
for the given digital system and put forward an intuitive explanation
of Benford's law by revealing the heterogeneity of such functions among
different first digits.
In Sec.~\ref{diftrans} we apply the Laplace transform to the digital indicator
functions to reveal their elegant properties.
From these, in Sec.~\ref{deriva} we provide a general derivation of a strict
version of Benford's law and prove that the strict Benford's law is composed
of a Benford term and an error term.
In Sec.~\ref{errorsec} we study the error term by applying our general result
to four categories of number sets, which obey Benford's law to varying degrees.
Especially, we prove that completely monotonic distributions can satisfy
Benford's law well.
Numerical studies are also provided to verify our method.
Sec.~\ref{summary} is reserved for conclusions.

\section{The intuition}\label{mantissa}
Let $F(x)$ be an arbitrary normalized probability density function (PDF) defined on the positive real number set $\mathbb{R}^+$ (here we use the capital letter $ F $ instead of the lowercase one, due to conventions for the Laplace transform introduced in Sec.~\ref{deriva}).
It does not matter if negative data are allowed, for we can instead use the PDFs of their absolute values.

%Considering
In the decimal system,
the probability $P_d$ of finding a number with first digit $d$ is the sum of the probability that it is within the interval $[d \cdotn 10^n, (d+1) \cdotn 10^n)$ for an integer $n$, therefore $P_d$ can be expressed as
\begin{equation}\label{sum}
P_d = \sum_{n = -\infty}^{\infty} \int_{d \cdot 10^n}^{(d+1) \cdot
	10^n} F(x) \,\dd  x \,,
\end{equation}
which can also be rewritten as
\begin{equation}\label{sumDisForm}
P_d = \int_{0}^{\infty} F(x)g_d(x) \,\dd  x \,,
\end{equation}
where $g_d(x)$ is the digital indicator function (DIF),
indicating numbers with first digit $ d $ in the decimal system
(here the lowercase letter is used, also due to conventions of the Laplace transform). Using the notation of the Heaviside step function,
\begin{equation}
\eta(x)= \left\{
\begin{array}{ll}
1 ,  &   \text{if $ x \geq 0 $\,, } \\
0  ,   &  \text{if $ x<0 $}\,,
\end{array}
\right.
\end{equation}
we can write $g_d(x)$ as
\begin{equation}\label{gDefine}
g_d(x)=\sum_{n = -\infty}^{\infty}\left[\eta (x-d \cdotn 10^n)- \eta
(x-(d+1) \cdotn 10^n)\right].
\end{equation}

\begin{figure}[h]
	\begin{center}
		\includegraphics[width=.7\linewidth]{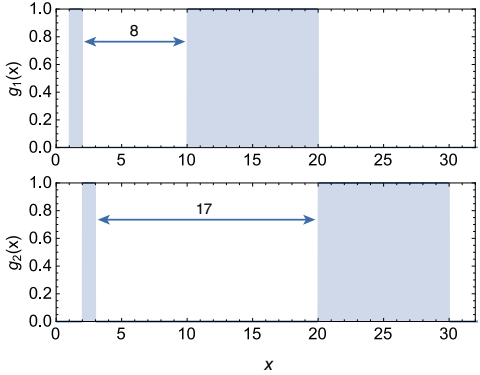}
		\caption{Images of digital indicator functions $g_1(x)$ and $g_2(x)$. Neither of them can be a translation or an expansion of the    other.}\label{illustration}
	\end{center}
\end{figure}

Different first digits define different $ g_d(x) $ functions, thus behave
differently in the digital system. For a better illustration, we draw the
images of $g_1(x)$ and $g_2(x)$ in the interval $[1,30)$, as shown in
Fig.~\ref{illustration}. We notice that $ g_2(x) $ can be neither a translation
nor an expansion of $ g_1(x) $, and that the gap between the shaded areas
in $ g_2(x) $ is wider than that in $ g_1(x) $. This fact intuitively explains
the inequality among the 9 digits, where smaller leading digits are more likely to appear.

Furthermore, if drawn on the logarithmic scale, $ g_d(x) $ becomes a
periodic function with a mean value of $ \log_{10}(1+\frac{1}{d}) $.
This gives us the intuition why $ g_d(x) $ has a strong connection
with Benford's law. In the following sections, through strict mathematical derivations,
we verify our intuition and show that the Benford term comes exactly from $ g_d(x) $.

\section{The Laplace transform of the digital indicator function}\label{diftrans}

In this section, we study the Laplace transform of the digital
indicator function~(DIF) and show that the transformed DIF
is also a log-periodic function which frequently appears in
various systems~\cite{sornette1998discrete}, and exhibits some
elegant properties that indicate Benford's law. For general cases,
we can define the DIF under base-$ b $ as $ g_{b,d,l}(x) $, whose
value is 1 for numbers within the interval $[d \cdotn b^n, (d+l) \cdotn b^n)$ for some integer $ n $ and 0 otherwise, i.e.,
\begin{equation}\label{gDefineGeneral}
g_{b,d,l}(x)=\sum_{n = -\infty}^{\infty}\left[\eta (x-d \cdotn b^n)-
\eta (x-(d+l) \cdotn b^n)\right].
\end{equation}

The Laplace transform of this general DIF is defined as
\begin{equation}\label{transDef}
G_{b,d,l}(t)=\int_{0}^{\infty} g_{b,d,l}(x)e^{-tx} \,\dd  x\,.
\end{equation}

We turn to the logarithmic scale again and further define
\begin{equation}\label{Hfunc}
\begin{split}
H_{b,d,l}(t)&=tG_{b,d,l}(t)\,,\\
\widetilde{H}_{b,d,l}(s)&=H_{b,d,l}(e^s)\,.
\end{split}
\end{equation}

The properties of $ \widetilde{H}_{b,d,l}(s) $ are given as follows:
\begin{enumerate}
	\item $ \widetilde{H}_{b,d,l}(s) $ is periodic with period $ \ln b $;
	\item the mean value of $ \widetilde{H}_{b,d,l}(s) $ within any single period is $ \log_b(1+\frac{l}{d}) $.
\end{enumerate}

The first property is obvious by expanding $ \widetilde{H}_{b,d,l}(s) $ from Eqs.~\ref{transDef} and \ref{Hfunc}, i.e.,
\begin{equation}\label{Hexact}
\begin{split}
&\widetilde{H}_{b,d,l}(s) = \\
&\,\,\sum_{n=-\infty}^{\infty}
\bigg(\exp\big[{-d\cdotn b^{(n+\frac{s}{\ln b})}}\big]-\exp\big[{-(d+l)\cdotn b^{(n+\frac{s}{\ln b})}}\big]\bigg).
\end{split}
\end{equation}
For the second property, we have the mean value of $ \widetilde{H}_{b,d,l}(s) $ within $ [0,\,\ln b) $ as
\begin{align}\label{Hmean}
\big\langle \widetilde H_{b,d,l}(s)\big\rangle
&= \frac{1}{\ln b}\int_{0}^{\ln b} \widetilde H_{b,d,l}(s) \,\dd s \nonumber\\
&= \frac{1}{\ln b}\int_{-\infty}^{\infty}\!\! \bigg(\!\! \exp[{-d\cdotn e^s}]-\exp[{-(d+l)\cdotn e^s}]\!\bigg)\dd s \nonumber\\
&= \frac{1}{\ln b}\int_{0}^{\infty}\frac{1}{t} \left( e^{-dt}-e^{-(d+l)t}\right) \,\dd t \nonumber\\
&= \log_b(1+\frac{l}{d})\,.
\end{align}

With these two properties, it is straightforward to rewrite $ \widetilde{H}_{b,d,l}(s) $ as
\begin{equation}\label{Deltadefine}
\widetilde H_{b,d,l}(s) =  \log_{b}(1+\frac{l}{d})+\widetilde \Delta_{b,d,l}(s)\,,
\end{equation}
where $ \widetilde \Delta_{b,d,l}(s) $ represents the periodic fluctuation of $ \widetilde{H}_{b,d,l}(s) $ around its mean value.
It is noted here that $ \widetilde{H}_{b,d,l}(s) $ is the logarithmic Laplace spectrum of the DIF.
Therefore, it is independent of any particular distributions of number sets.

The first term $ \log_{b}(1+\frac{l}{d}) $ is here called the Benford term and we will
show in Sec.~\ref{deriva} that it is the origin of the classical Benford's law,
while $ \widetilde \Delta_{b,d,l}(s) $ is responsible for the possible deviation
from the law. We will show in Sec.\ref{errorsec} that this deviation is small for a
big family of distributions.

It is worth noting that the Benford term is derived merely from the DIF of a certain digital
system without assuming the exact form of the PDF. Therefore, we assert that the origin of
Benford's law comes from the way that the digital system is constructed, instead of the way that some
specific number set is formed.

\section{The derivation of the general digit law}\label{deriva}

We see that the logarithmic Laplace spectrum of the digital indicator function fluctuates
around the Benford term. Another reason why we choose the Laplace transform is that the inverse
Laplace transform can be served as a method to judge how well a specific PDF obeys the law, as well as to derive the general digit law.
For an arbitrary PDF $ F(x) $, we can assume that it has an inverse Laplace transform $ f(t) $ which belongs to $ L^1(\mathbb{R}^+) $, satisfying
\begin{equation}\label{tansDef}
F(x)=\int_{0}^{\infty} f(t)e^{-tx} \,\dd  t\,.
\end{equation}

The probability that a number drawn from a data set with a PDF $ F(x) $ is within the set $ \bigcup_{n=-\infty}^\infty [d,\,d+l) \times b^n $ can be expressed as
\begin{equation}\label{PGen}
P_{b,d,l} = \int_{0}^{\infty} F(x)g_{b,d,l}(x) \,\dd  x \,.
\end{equation}
We turn to the logarithmic scale again and define
\begin{equation}
\widetilde f(s) =f(e^s)\,,
\end{equation}
then $ \widetilde f(s) $ also satisfies the normalization condition, i.e.,
\begin{equation}\label{norm}
\int_{-\infty}^{\infty}\widetilde{f}(s) \,\dd  s\
= \int_{0}^{\infty}\frac{f(t)}{t}\,\dd t =
\int_{0}^{\infty}F(x)\,\dd x = 1\,.
\end{equation}
According to the property of the Laplace transform, Eq.~\ref{PGen} can be rewritten in the inverse Laplace space of the PDF as
\begin{align}\label{tansChange}
\int_{0}^{\infty} F(x)g_{b,d,l}(x) \,\dd  x
%    &= \int_{0}^{\infty}\,\dd  x\,
%    g_{b,d,l}(x) \int_{0}^{\infty} f(t)e^{-tx}\,\dd t \nonumber\\
%    &= \int_{0}^{\infty}\,\dd  t\, f(t) \int_{0}^{\infty} g_{b,d,l}(x)e^{-tx}
%    \,\dd x \nonumber \\
&= \int_{0}^{\infty} f(t)G_{b,d,l}(t) \,\dd  t \nonumber\\
&= \int_{-\infty}^{\infty} \widetilde f(s) \widetilde H_{b,d,l}(s) \,\dd  s \,.
\end{align}

Combining the expression of $ \widetilde H_{b,d,l}(s) $ in Eq.~\ref{Deltadefine} and the normalization condition of $ \widetilde f(s) $ in Eq.~\ref{norm}, we derive the strict form of Benford's law, which is composed of a Benford term and an error term, as follows,
\begin{equation}\label{Pfin}
P_{b,d,l}=\log_b(1+\frac{l}{d})+\int_{-\infty}^{\infty} \widetilde f(s) \widetilde \Delta_{b,d,l}(s) \,\dd  s \,.
\end{equation}

Since $ \widetilde \Delta_{b,d,l}(s) $ is slightly fluctuating, the error term is small for most circumstances, as we intend to illustrate further in Sec.~\ref{errorsec}. If we ignore the error term in Eq.~\ref{Pfin}, the strict Benford's law turns into the general digit law, i.e.,
\begin{equation}
P_{b,d,l}\approx \log_b(1+\frac{l}{d})\,.
\end{equation}

Lots of variations of the classical Benford's law can be seen as corollaries of the general digit law. For example, the base $ b $ can be set to 100 to derive the second significant digit law given by Newcomb~\cite{n81}. A number $ (x)_{10} $ in the decimal system can be equally treated as a number $ (x)_{100} $ in the base-100 system, so that the second digit of $ (x)_{10} $ being $ d $ is equivalent to that either the first ``digit'' of $ (x)_{100} $ belongs to the set $ S_d=\{10+d,\,20+d,\cdots,\,90+d\} $, or that the first ``digit'' of $ (10x)_{100} $ belongs to the same set $ S_d $. Therefore, we have
\begin{align}
	 P\big(&\text{2nd digit of $ (x)_{10} $}=d\big) \nonumber\\
	&=P\big(\text{1st ``digit'' of $ (x)_{100}\in S_d $}\big) \,+ \nonumber\\
	&\qquad\qquad\qquad P\big(\text{1st ``digit'' of $ (10x)_{100} \in S_d $}\big) \nonumber\\
	&\approx\sum_{k=1}^9\log_{100} \left(1+\frac{1}{10k+d}\right) \!+\! \sum_{k=1}^9\log_{100}\left(1+\frac{1}{10k+d}\right) \nonumber\\
	&=\sum_{k=1}^9\log_{10}\left(1+\frac{1}{10k+d}\right).
\end{align}

Similar reasoning can also be applied to the $ i $th-significant digit law of Hill~\cite{h95c}:
letting $D_i$ ($D_1,D_2,...$) denotes the $i$th-significant digit (with base 10) of a number (e.g., $D_1(0.0314) = 3$, $D_2(0.0314)=1$, $D_3(0.0314) = 4$), then for all positive integers $k$ and all $d_j \in {0,1,\cdots,9}$, $j=1,2,\cdots,k$, one has
\begin{equation}\label{hill}
\begin{split}
P(D_1=d_1,\,\cdots,\,&D_k=d_k)\approx\\
&\log_{10}\left[1+\left(\sum_{i=1}^k d_i \cdotn
10^{k-i}\right)^{-1}\right].
\end{split}
\end{equation}

%==========================================

\section{The error term}\label{errorsec}

\begin{figure}[h]
	\centering
	\includegraphics[width=.7\linewidth]{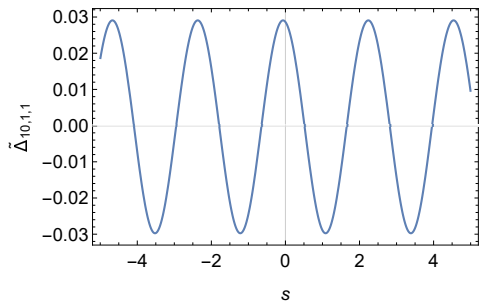}
	\caption{The image of the $ \ln 10 $-periodic function $ \widetilde{\Delta}_{10,1,1}(s) $ for the first digit 1 in the decimal system.}
	\label{delta10}
\end{figure}

In this section, we introduce a method to judge how well a certain PDF obeys the classical Benford's law by analyzing the total error term in Eq.~\ref{Pfin}, which is the interrelation of $ \widetilde \Delta_{b,d,l}(s) $ and $ \widetilde f(s) $, i.e.,
\begin{equation}\label{totalError}
\Delta_{\mathrm{total},b,d,l}=\int_{-\infty}^{\infty} \widetilde f(s) \widetilde \Delta_{b,d,l}(s) \,\dd  s \,.
\end{equation}
We know in Sec.~\ref{deriva} that $ \widetilde \Delta_{b,d,l}(s) $ is a $ \ln b $-periodic function with a mean value of 0. For instance, a graph of $ \widetilde \Delta_{10,1,1}(s) $ is shown in Fig.~\ref{delta10}. The amplitude of this periodic function is small compared with the Benford term, e.g., the amplitude of $ \widetilde \Delta_{10,1,1}(s) $ is less than 0.03 while the Benford term is 0.30.
Therefore, intuitively speaking, if $ \widetilde f(s) $ is smooth enough and changes slowly, its interrelation with $ \widetilde \Delta_{b,d,l}(s) $ tends to be averaged out; thus the total error tends to be small. On the other hand, if $ \widetilde f(s) $ oscillates violently, the interrelation is highly sensitive to the exact form of $ \widetilde f(s) $, and the total error is likely to be large.

Rigorously, we can classify real-world number sets into the following four categories,
with different degrees of deviation from the classical Benford's law.
\begin{enumerate}
	\item The error term equals to the constant 0 for scale-invariant distributions.
	\item Since the error term is bounded by a proportion of the $ L^1 $-norm
	of $ \widetilde f(s) $ (noted by $ \|\widetilde f \|_1 $), i.e.,
	\begin{equation}\label{Mcondition}
	\Delta_{\mathrm{total},b,d,l}\in \left[\|\widetilde f \|_1 \min\{\widetilde\Delta_{b,d,l}\},\,
	\|\widetilde f \|_1 \max\{\widetilde\Delta_{b,d,l}\} \right],
	\end{equation}
	when $ \widetilde{f}(s) $ oscillates
	mildly between positive and negative values, $ \|\widetilde f \|_1 $ is close to 1,
	thus the error term is small.
	\item Specifically, if $ \widetilde{f}(s) \geq 0 $ holds for $ \forall s\in \mathbb{R} $,
	$ \|\widetilde f \|_1 $ reaches its minimum value 1, so the bound is the tightest, i.e.,
	\begin{equation}\label{M1condition}
	\Delta_{\mathrm{total},b,d,l}\in \left[ \min\{\widetilde\Delta_{b,d,l}\},\,
	\max\{\widetilde\Delta_{b,d,l}\} \right].
	\end{equation}
	Such distributions are called completely monotonic distributions.
	\item When $ \widetilde f(s) $ oscillates dramatically, its $ L^1 $-norm becomes large, so the error term becomes uncertain and the classical Benford's law is generally violated.
	
\end{enumerate}
We prove and explain the above assertions one by one in the following sections. Some numerical examples are provided for better illustration.

\subsection{Scale-invariant distribution}
We first turn to scale-invariant distributions which are initially discussed by Hill
on the prospect of the probability measure theory~\cite{h95b}.  Pietronero {\it et al.}~\cite{pietronero2001explaining}
has shown that scale invariant distributions arise naturally from any multiplicative
stochastic process such as the dynamics of stock prices. With Laplace transform, we can also
show that scale invariance leads to Benford's law.

Following the definition of Hill \cite{h95b}, a scale-invariant probability measure $ P $ is a measure defined
on the following $ \sigma $-algebra
\begin{equation}\label{scaleInv}
\mathscr{M}=\bigg\lbrace S\;\bigg|\; S=\bigcup_{n=-\infty}^{\infty} B\times b^n \bigg\rbrace  ,\;
\text{for some Borel }  B\subseteq [1,b),
\end{equation}
satisfying $ P(S)=P(\lambda  S) $ for all $ \lambda>0 $ and $ S\in \mathscr{M} $. Hill proved that such a scale-invariant measure strictly satisfies Benford's law. We can easily prove this result again with the language of PDF, if we set $ S $ to be $ \bigcup_{n=-\infty}^{\infty} [d,\,d+l)\times b^n $ and notice that the scale-invariance property implies the following statement, i.e, for all $ \epsilon\in \mathbb{R} $,
\begin{equation}\label{hillep}
\int_{-\infty}^{\infty} \widetilde f(s)\widetilde\Delta_{b,d,l}(s)\,\dd s=\int_{-\infty}^{\infty}\widetilde f(s)\widetilde \Delta_{b,d,l}(s+\epsilon)\,\dd s=C
\end{equation}
holds, where $ C $ is independent of $ \epsilon $.
According to the periodic and zero-mean properties of $ \widetilde \Delta_{b,d,l}(s) $, we have
\begin{align}
\int_{0}^{\ln b}\bigg(\int_{-\infty}^{\infty}&\widetilde f(s)\widetilde \Delta_{b,d,l}(s+\epsilon)\,\dd s \bigg) \,\dd \epsilon \nonumber\\
&= \int_{-\infty}^{\infty}\widetilde f(s)\left(  \int_{0}^{\ln b} \widetilde \Delta_{b,d,l}(s+\epsilon)\,\dd \epsilon \right)  \,\dd s \nonumber\\
&=\int_{-\infty}^{\infty}\widetilde f(s)\,0\cdot \dd s \nonumber\\
&=0\,.
\end{align}
Thus, we get $ \int_{0}^{\ln b}C\,\dd s=0 $, i.e., $ C=0 $. We notice that $ C $ is also the error term in Eq.~\ref{totalError}. Therefore, such scale-invariant distributions conform strictly to the classical Benford's law.

\subsection{Small-$ \|\widetilde f \|_1 $ distribution}

Although scale invariance is common in nature, not all natural data sets are scale invariant.
Even for those data sets which are not scale invariant, Benford's law can still be a good approximation for most cases.
This is because the error term in Eq.~\ref{totalError} can be well bounded by the $ L^1 $-norm
of $ \widetilde f $, as is shown in Eq.~\ref{Mcondition}.

The proof is from the fact that $ \|\widetilde f \|_1 $ is an upper bound of the integral of a function.
According to Eq.~\ref{totalError}, we have
\begin{equation}
\begin{split}
\|\widetilde f \|_1 \min \{ \widetilde \Delta_{b,d,l}\}
& = \min\{\widetilde \Delta_{b,d,l}\}\int_{-\infty}^{\infty}\abs{\widetilde{f}(s)}\,\dd s \\
& \leq \Delta_{\mathrm{total},b,d,l} \\
& \leq \max\{\widetilde \Delta_{b,d,l}\}\int_{-\infty}^{\infty}\abs{\widetilde{f}(s)}\,\dd s \, \\
& = \|\widetilde f \|_1 \max \{ \widetilde \Delta_{b,d,l}\} ,
\end{split}
\end{equation}
where $ \|\widetilde f \|_1 = \int_{-\infty}^{\infty}\abs{\widetilde{f}(s)}\,\dd s $ is the $ L^1 $-norm of $ \widetilde f(s) $.

In the decimal system, we numerically calculate the Benford terms (noted by $ P^\mathrm{B}_{10,d,1} $) and
the maximum value of $\abs{\widetilde \Delta_{b,d,l}}$ (noted by $ \Delta^\mathrm{max}_{10,d,1} $) in Table~\ref{diffDigit}.
The relative errors $ \delta_{10,d,1}^{\mathrm{max}}=\Delta^\mathrm{max}_{10,d,1}/P^\mathrm{B}_{10,d,1} $ are also listed.
From Table~\ref{diffDigit}, we notice that when $ \|\widetilde f \|_1 $ is small,
Benford's law holds well, with a maximum relative error
of $ 12* \|\widetilde f \|_1 \% $ for all digits.

\begin{table*}[htbp]
	\centering
	\caption{Numerical results of the Benford term $ P^\mathrm{B}_{10,d,1} $ and the maximum of the
	absolute error term $ \max\{\abs{\widetilde \Delta_{b,d,l}}\} $  in the decimal system
	, together with the relative errors $ \delta_{10,d,1}^{\mathrm{max}} $.}
	\begin{tabular}{cccccccccc}
		\hline\hline
		$ d $    & 1    & 2    & 3    & 4    & 5    & 6    & 7    & 8    & 9 \\
		\hline
		$ P^\mathrm{B}_{10,d,1}\,/\%$    & 30.10 & 17.61 & 12.49 & 9.69 & 7.92 & 6.69 & 5.80  & 5.12 & 4.58 \\
		$ \Delta^\mathrm{max}_{10,d,1}\,/\% $    & 2.97 & 1.94 & 1.41 & 1.11 & 0.91 & 0.76 & 0.77 & 0.59 & 0.53 \\
		$ \delta_{10,d,1}^{\mathrm{max}}\,/\%  $ & 9.9  & 11.0 & 11.3 & 11.4 & 11.5 & 11.5 & 11.6 & 11.6 & 11.7 \\
		\hline\hline
	\end{tabular}%
\label{diffDigit}%
\end{table*}%

The exponential distribution is a good example of small-$ \|\widetilde f \|_1 $ distributions.
\begin{equation}\label{exp}
F(x)=\lambda e^{-\lambda x}\;(\lambda>0)\,.
\end{equation}
Engel and Leuenberger showed that the exponential distribution obeys Benford's law approximately within bounds of 0.03 (for $ b=10 $ and $ d=l=1 $)~\cite{el03}.
This fact can be explained by Eq.~\ref{M1condition} if we notice that the logarithmic inverse Laplace transform
of the exponential distribution is $ \widetilde f(s) = \delta(s-\ln \lambda) $. Therefore, $ \|\widetilde f \|_1 =1$, so
\begin{equation}
\begin{split}
\Delta_{\mathrm{total},10,1,1}&\in \left[\min\{\widetilde\Delta_{10,1,1}\},\,\max\{\widetilde\Delta_{10,1,1}\}\right]\\
&\subset (-0.03,\,0.03) \,.
\end{split}
\end{equation}

The log-normal distribution with a big variance is another example. The PDF is
\begin{equation}\label{loguni}
F(x)=\frac{1}{x\sigma\sqrt{2\pi}}e^{-\frac{(\ln x-\mu)^2}{2\sigma^2}}\,.
\end{equation}
As long as $ \sigma $ is not too small relative to the base $ b $, $ \widetilde f(s) $ oscillates mildly between positive and negative values, so $ \|\widetilde{f}\|_1 $ is also considerably small.
For example, when $ \mu=5 $ and $ \sigma=1 $,
\begin{equation}
\|\widetilde f \|_1= \int_{-\infty}^{\infty}\abs{\widetilde{f}(s)}\,\dd s =1.610\,.
\end{equation}
Then, from Eq.~\ref{Mcondition} we have
\begin{equation}
\begin{split}
\Delta_{\mathrm{total},10,1,1}
&\in \Big[\|\widetilde f \|_1 \min\{\widetilde\Delta_{10,1,1}\},\,
\|\widetilde f \|_1 \max\{\widetilde\Delta_{10,1,1}\} \Big] \\
&\subset (-0.048,\,0.047)\,,
\end{split}
\end{equation}
which is also acceptable compared to the Benford term 0.301.

\subsection{Completely monotonic distribution}

The family of completely monotonic (c.m.) distributions is
a special case of small-$ \|\widetilde f \|_1 $ distributions.
Completely monotonic distributions are probability distributions with c.m. PDFs.
This is equivalent to say that $ \widetilde f(s) $ is non-negative for all $ s\in\mathbb{R} $.
In this case, $ \|\widetilde f \|_1 $ reaches its minimum value 1 due to the
normalization condition in Eq.~\ref{norm}, thus
we get the tightest bound in Eq.~\ref{M1condition}.
In fact, the exponential distributions and
the scale invariant distributions that we have discussed above
are both c.m., but the family of c.m. functions are much more prosperous.
Miller and Samko~\cite{miller2001completely} surveyed a series of good properties of completely
monotonic functions.
Herein we summarize these properties again for the convenience of the readers.
\begin{enumerate}
	\item A function $F$ with domain $(0,\infty)$ is said to be c.m. if it
	possesses derivatives $F^{(n)}(x)$ for all $n=0,1,2,3,...$ and if $(-1)^{n}F^{(n)}(x)\geq 0$ for all $x>0$.
	\item (Bernstein-Widder theorem) $F(x)$ is c.m. if and only if
	$F(x)$ is the Laplace transform of a non-negative measurable function $f(t)\colon t \to [0,+\infty)$ and
	$F(x)<+\infty$ for $0<x<+\infty$.
	\item \label{elementary} The following elementary functions are all c.m. functions:
	\begin{equation}
		\begin{aligned}
		\begin{split}
		 e^{-a x}, \quad &  a \geq 0, \\
		 \frac{1}{(a+c x)^{\alpha}},  \quad & a,c,\alpha \geq 0 , \\
		 \ln (a+\frac{c}{x}), \quad & a \geq 1, c>0. \\
		\end{split}
		\end{aligned}
	\end{equation}	
	\item \label{gen1} If $F(x)$ is c.m., then $e^{F(x)}$ is also c.m.
	\item \label{gen2} If $F_{1}(x)$ and $F_{2}(x)$ are c.m., then $ a F_{1}(x)+c F_{2}(x),\, a\geq 0,\, c\geq 0 $
	is also c.m.
	\item \label{gen3} If $F_{1}(x)$ and $F_{2}(x)$ are c.m., then $F_{1}(x)F_{2}(x)$ is also c.m.
	\item \label{gen4} Let $F(x)$ be c.m. and let $\tau(x)$ be nonnegative with a c.m. derivative,
	then $F(\tau(x))$ is also c.m.

\end{enumerate}

From properties \ref{gen1}, \ref{gen2}, \ref{gen3}, \ref{gen4}, we can generate a large family of c.m. functions
from elementary c.m. functions in Property \ref{elementary}.
For such a c.m. function $F(x)$ to be a valid PDF,
it should also satisfy the normalization condition in Eq.~\ref{norm}.
When
\begin{equation} \label{normlessthaninfinity}
	\int_{0}^{\infty} F(x) \, \dd  x < \infty,	
\end{equation}
the normalization condition can be
guaranteed by introducing a normalization factor.

Several examples of c.m. PDFs are listed below. The parameters are thus chosen so that
the integral in Eq.~\ref{normlessthaninfinity} converges.
The normalization factors are omitted in Eq.~\ref{cmpdf}.
Distributions generated from these PDFs all satisfy Benford's law within a very small bound:
	\begin{equation} \label{cmpdf}
		\begin{aligned}
		\begin{split}
		 e^{-a(x+c)^{\alpha}}, \quad & \; a,c \geq 0, \; 0 \leq \alpha \leq 1 ,\\
		 e^{a(x+c)^{\alpha}}-1, \quad & \; a, c > 0, \;  \alpha < 0 ,\\
		\frac{1}{(x+c)^{\alpha}}, \quad &\; c> 0,\; \alpha > 1 ,\\
		\frac{1}{x^{\nu}} e^{-ax^{\alpha}}, \quad &\; a\geq 0,\; 0\leq \alpha \leq 1, \; 0\leq \nu <1 ,\\
		e^{-a(\ln x+c)^{\alpha}}, \quad & \; a,c \geq 0, \; 0 \leq \alpha \leq 1 .\\
		\end{split}
		\end{aligned}
	\end{equation}	

Some non-c.m. distributions can be converted to c.m. distributions through a
non-linear transformation of the data. Literature has shown that non-linear transformations
on some data sets yield more robust results when
Benford's law is used to detect fraud \cite{clippe2012benford}.
To explain this, suppose $x$ is a random
variable with PDF $F(x)$, if we transform $x$ into $y=\tau(x)$, then the PDF of $y$ becomes
\begin{equation} \label{transformpdf}
	\sum _{k=1}^{n(y)}\frac{ F(\tau_{k}^{-1}(y))}{\abs{\tau_{k}^{\prime}(\tau_{k}^{-1}(y))}},
\end{equation}
where $n(y)$ is the number of solutions in $x$ for the equation $\tau(x)=y$ and $\tau_{k}^{-1}(y)$
is the $k$th solution.

One example of this case is the normal distribution with the PDF
\begin{equation}\label{normal}
	 F(x) = \frac{1}{\sqrt{2\pi\sigma^{2}}}e^{-\frac{(x-\mu)^{2}}{2\sigma^{2}}} .
\end{equation}
Through the transformation $y=(x-\mu)^{2}$, the PDF becomes
\begin{equation} \label{transformnormal}
	F(y)=\frac{1}{\sqrt{2\pi\sigma^{2}y}} e^{-\frac{y}{ 2\sigma^{2} }} , \quad y>0 .
\end{equation}
Eq.~\ref{transformnormal} is c.m. Therefore, the transformed data set of $y$ fulfills
Benford's law with an error bound of 0.03, same to that of exponential distributions.

\subsection{Violently-oscillating-$ \widetilde{f}(s) $ distribution}
The only case in which Benford's law loses its power is when $ \widetilde{f}(s) $ oscillates violently
between positive and negative values. The fast oscillation of $ \widetilde{f}(s) $ makes the small term of
$ \widetilde \Delta_{b,d,l} $ be counted and accumulated again and again. Hence,
$ \|\widetilde f \|_1 $ becomes large, and $ \widetilde{f}(s) $ is highly sensitive to some tiny
perturbation on $ F(x) $, reflecting the instability of the inverse Laplace transform~\cite{bellman1966numerical}.
Therefore, the total error is also highly sensitive to the exact form of $ F(x) $ and
Benford's law is generally violated in this case.

\begin{figure}[h]
	\centering
	\includegraphics[width=.7\linewidth]{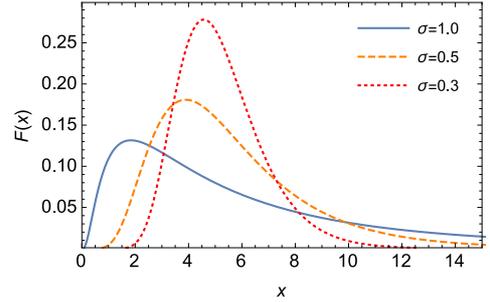}%dash
	\caption{The images of log-normal PDF $ F(x) $ with $ \sigma=1.0 $, $ 0.5 $ and $ 0.3 $, with $ \mu=\ln 5 $ fixed.}
	\label{lognorm_prob}
\end{figure}
\begin{figure}[!t]
	\centering
	\includegraphics[width=.7\linewidth]{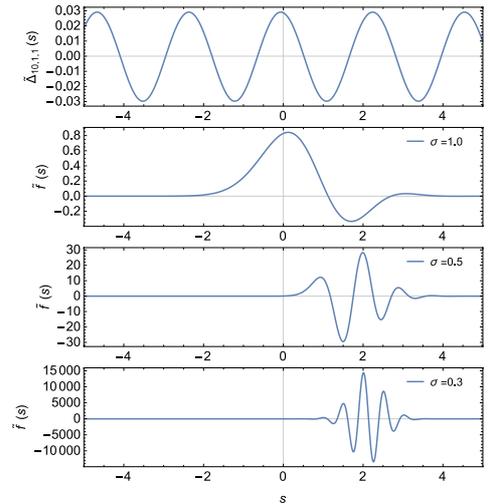}
	\caption{The images of $ \widetilde \Delta_{10,1,1}(s) $ together with $ \widetilde f(s) $ for the log-normal distribution with $ \sigma=1.0 $, $ 0.5 $ and $ 0.3 $. Noted that $ \widetilde{f}(s) $ displays stronger oscillatory behavior when $ \sigma $ decreases.}
	\label{ltlog}
\end{figure}
\begin{figure}[!t]
	\centering
	\includegraphics[width=.7\linewidth]{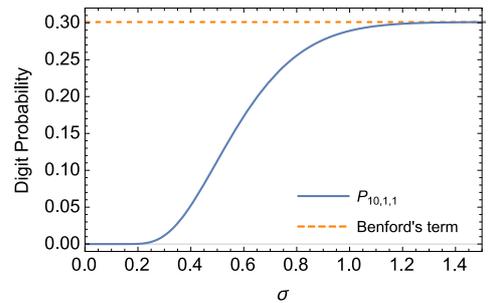}%dash
	\caption{$ P_{10,1,1} $ of the log-normal distribution compared with the Benford term $ P^B_{10,1,1} $ under $ 0<\sigma \leq 1.5 $.}
	\label{lognorm_curve}
\end{figure}

Examples of such violently-oscillating-$ \widetilde f(s) $ distributions are log-normal distributions with small $ \sigma $ and uniform distributions.
We have shown that the log-normal distribution with parameters $ \mu=\ln 5 $ and $ \sigma=1.0 $ approximately conforms to the classical Benford's law. However, when $ \sigma $ becomes smaller, the distribution is concentrated on some specific first digits, as shown in Fig.~\ref{lognorm_prob} for $ \sigma=1.0$, 0.5, 0.3. In the logarithmic inverse Laplace space, we numerically calculate $ \widetilde{f}(s) $ by the Stehfest method~\cite{steh1,steh2,steh3} and plot them together with $ \widetilde\Delta_{10,1,1}(s) $ in Fig.~\ref{ltlog}.
We notice that $ \widetilde{f}(s) $ displays stronger oscillatory behavior as $ \sigma $ decreases. Thus the interrelation between $ \widetilde\Delta_{10,1,1}(s) $ and $ \widetilde f(s) $ is highly sensitive to the exact form of $ \widetilde f(s) $, or some tiny perturbation on $ F(x) $.

In fact, we can numerically calculate $ P_{10,1,1} $ directly from Eq.~\ref{sumDisForm} for $ 0<\sigma\leq1.5 $ and the results are shown in Fig.~\ref{lognorm_curve}.
This verifies our prediction. When $ \sigma $ becomes smaller, $ \widetilde f(s) $ oscillates stronger and $ P_{10,1,1} $ deviates further from the Benford term.

\begin{figure}[!t]
	\centering
	\includegraphics[width=.7\linewidth]{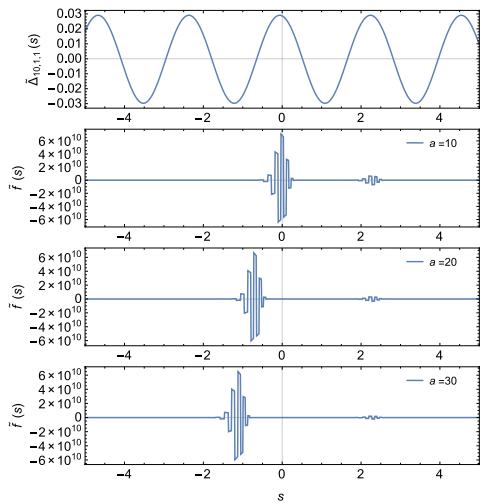}
	\caption{The images of $ \widetilde \Delta_{10,1,1}(s) $ together with $ \widetilde f(s) $ for the uniform distribution with $ a=10 $, $ 20 $ and $ 30 $. Be noted that $ \widetilde{f}(s) $ functions oscillate much more fiercely than those in Fig.~\ref{ltlog}.}
	\label{u0}
\end{figure}

%%=======
For the uniform distribution on the interval $ [1,\,a]\;(a>1) $, $ \widetilde f(s) $ oscillates even stronger. The PDF is given by
\begin{equation}
F(x)=\frac{\eta(x-1)-\eta(x-a)}{a-1}\,.
\end{equation}
We use an analytic function to approach $ F(x) $, and calculate the numerical values of the inverse Laplace transform for $ a = 10 $, 20 and 30, as shown in Fig.~\ref{u0}.
Since $ \widetilde f(s) $ is extremely unstable for such distributions, we can expect the total error is unstable as well, generally large.

\begin{figure}[!t]
	\centering
	\includegraphics[width=.7\linewidth]{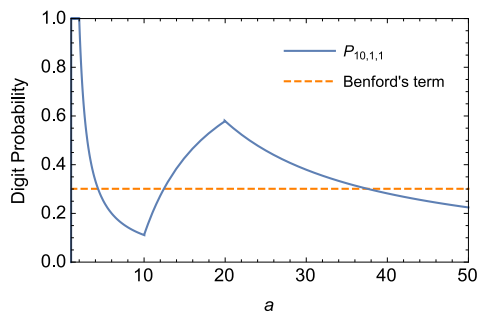}%dash
	\caption{$ P_{10,1,1} $ of the uniform distribution compared with the Benford term $ P^B_{10,1,1} $ under $ 0<a\leq 50 $.}
	\label{uni_curve}
\end{figure}

Also for verification, we draw the numerical values of $ P_{10,1,1} $ under $ 1<a\leq 50 $ in Fig.~\ref{uni_curve}. $ P_{10,1,1} $  in this case depends greatly on the endpoints of the PDFs, occasionally coincides with the Benford term but generally violates the classical Benford's law. This is again expected.

At last, we want to rectify two typical misunderstandings about Benford's law,
i.e.,
\begin{enumerate}
\item random data sets without human manipulation are supposed to fulfill Benford's law;
\item smooth distributions that span many orders of magnitude should satisfy Benford's law.
\end{enumerate}
Unfortunately, however, neither of these two assertions are correct.

As we have shown, the first statement is only approximately true for random variables generated from
PDFs with small $ \|\widetilde f \|_1 $. Even natural random data sets could violate Benford's law
if their PDFs oscillate violently in the inverse Laplace space. For example, lots of
natural number sets are distributed normally or lognormally near their mean values with
very small variances, such as heights of all trees on the earth.
Such data sets, although natural, do not obey Benford's law.

As for the second statement, whether or not a distribution satisfies
Benford's law is determined by the shape, instead of the scale, of the PDF.
Therefore, even an extremely flat PDF which spans many orders of magnitude
may still violate Benford's law. To understand this, we note that one can
change the scale of any PDF $F_{1}(x)$ by
multiplying the original data with an arbitrary number $a$.
The PDF of the new data set is
\begin{equation} \label{scalechange}
	F_{2}(x)=\frac{1}{a}F_{1}(\frac{x}{a}). 	
\end{equation}
When $a$ turns bigger, $F_{2}(x)$ is flattened out, and it can span as
many orders of magnitude as we desire.
However, the logarithmic inverse Laplace transform of $F_{1}(x)$
and $F_{2}(x)$ differ only by a horizontal shift, i.e.,
\begin{equation} \label{horizontalshift}
	\widetilde f_{2}(s)=\widetilde f_{1}(s+\ln(a)) .	
\end{equation}
Such a horizontal shift does not change the unstable nature of the error
term in Eq.~\ref{totalError}.

If we desire to reduce the error term, we need to flatten out
$\widetilde f_{1}(s)$ directly, e.g.,
into $\widetilde f_{2}(s) = \frac{1}{\alpha}\widetilde f_{1}(\frac{s}{\alpha})$.
When $\alpha$ becomes bigger, the total error of Benford's law in Eq.~\ref{totalError},
as the interrelation between a periodic function and an extremely flat
$\widetilde f_{2}(s)$, tends to vanish.
In this case, the shape of the PDF $F_{1}(x)$ has been changed.
In fact, when $\alpha$ is big enough and $f_{1} (1) \neq 0$ (this can be guaranteed up to
a scaling factor in Eq.~\ref{scalechange}), $F_{2}(x)$ approaches to the scale invariant distribution, i.e.,
\begin{equation}
	\begin{split}
	\frac{F_{2}(x)}{F_{2}(cx)}
	&= \frac{\int_{0}^{\infty} \frac{1}{\alpha}f_{1}(t^{\frac{1}{\alpha}})e^{-tx} \,\dd  t\, }
	{\int_{0}^{\infty} \frac{1}{\alpha}f_{1}(t^{\frac{1}{\alpha}})e^{-ctx} \,\dd  t\, } \\
	&= \frac{\int_{0}^{\infty} f_{1}(t^{\frac{1}{\alpha}})e^{-tx} \,\dd  t\, }
	{\frac{1}{c} \int_{0}^{\infty} f_{1}((\frac{t}{c} )^{\frac{1}{\alpha}})e^{-tx} \,\dd  t\, } \\	
	& \xrightarrow[\text{pointwise}]{\alpha \to +\infty}	c, \quad \text{for} \; \forall c>0 .
	\end{split}
\end{equation}
Such an operation brings Benford's law back to power again because
the shape of the original PDF, not only the scale, has been changed.

%==========================================

\section{Summary}\label{summary}

The first digit law has revealed an astonishing regularity of
natural number sets. We introduce a method of the Laplace
transform to study the law in depth. Our method can explain the
long-standing puzzle about Benford's law, i.e., whether or not
Benford's law is merely a result of the way of writing numbers.
Our answer is yes in the sense that the Benford term can be
derived independently of any specific probability distributions.
This does not conflict with the fact that when the $ L^1 $-norm of
the logarithmic inverse Laplace transform of the PDF is large, Benford's law is
always violated.

Besides, the method sets a bound on the error term, allowing us
to predict the validity of Benford's law by the logarithmic inverse Laplace
transform of an arbitrary PDF. Real-world distributions can be
categorized into four types, corresponding to their oscillatory
behavior in the inverse Laplace space. A milder oscillation
guarantees higher conformity to the law, and vice versa. Especially,
the whole family of completely monotonic distributions all obey
Benford's law within a small bound. Numerical examples are shown
to verify our method. It is not strange anymore why Benford's law
is so successful in various domains of human knowledge.
Such a law should receive attention as a basic mathematical knowledge,
with great potential for vast application.

\section*{Acknowledgments}

This work is supported by National Natural Science Foundation of China (Grant No.~11475006). It is also supported by National Innovation Training Program for Undergraduates.

%\section*{References}

\bibliography{ref2}

\end{document}